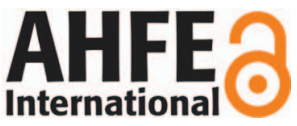

# A Lexical Analysis of Online Reviews on Human-AI Interactions

**Parisa Arbab and Xiaowen Fang**

DePaul University Chicago, IL 60604, USA

## ABSTRACT

This study focuses on understanding the complex dynamics between humans and AI systems by analyzing user reviews. While previous research has explored various aspects of human-AI interaction, such as user perceptions and ethical considerations, there remains a gap in understanding the specific concerns and challenges users face. By using a lexical approach to analyze 55,968 online reviews from G2.com, Producthunt.com, and Trustpilot.com, this preliminary research aims to analyze human-AI interaction. Initial results from factor analysis reveal key factors influencing these interactions. The study aims to provide deeper insights into these factors through content analysis, contributing to the development of more user-centric AI systems. The findings are expected to enhance our understanding of human-AI interaction and inform future AI technology and user experience improvements.

## INTRODUCTION

The integration of artificial intelligence (AI) tools into various domains has significantly impacted human-AI interaction dynamics. Understanding user perceptions, trust formation, ethical considerations, and societal impacts in the context of AI technologies is crucial for successful implementation and user acceptance. The literature review delves into the complexities of human-AI interaction, highlighting the importance of addressing user concerns and challenges to enhance the usability and effectiveness of AI systems. Despite advancements in AI technology, there exists a gap in understanding the specific concerns and challenges faced by users during their interactions with AI systems. User perceptions, trust formation, ethical considerations, and societal impacts play a critical role in shaping human-AI interaction dynamics. Addressing these factors is essential for ensuring positive user experiences and fostering trust in AI technologies. This research is important as it aims to bridge the gap in understanding user concerns and challenges in human-AI interaction. By analysing AI software reviews and employing a lexical approach, the study seeks to identify major issues encountered by users, contributing to a nuanced understanding of human-AI interaction dynamics. The findings are expected to provide valuable insights into the critical factors shaping human-AI interaction, informing future developments in AI technology and user experience.

## LITERATURE REVIEW

In recent years, the integration of artificial intelligence (AI) into various domains has sparked significant interest and raised numerous questions regarding its interaction with humans and its broader societal implications. This literature review aims to synthesize insights from five key sources to understand human-AI interaction and the societal impacts of AI technology.

Maadi et al. (2021) examined the interaction between humans and AI in machine learning, focusing on medical applications. The researchers conducted a literature review to understand the role of humans in AI processes, stages of interaction, individuals involved, and mechanisms of interaction. They highlighted the necessity of human involvement in AI processes, specific interaction stages, roles of individuals, and engagement methods. This study is relevant for understanding human-AI interaction intricacies, complementing user concerns in AI technology reviews.

Peltier et al. (2023) research aims to develop a conceptual framework for AI in interactive marketing and propose a research agenda. Through a literature review, the researchers synthesized existing theories and research on AI in marketing to create a framework for value co-creation. The resulting framework includes historical perspectives, AI definitions, impact on marketing, relevant theories, and synthesized research. It contributes insights into enhancing human-AI interaction dynamics in marketing contexts.

Asan et al. (2020) study focuses on clinicians and explores how AI impacts human trust in healthcare, specifically examining trust in service delivery. Likely involving surveys, interviews, or observations, the research highlights factors influencing user attitudes towards healthcare AI systems, including personal experiences and social reputation. It provides insights into trust dynamics between clinicians and AI technologies, crucial for the adoption and acceptance of AI in healthcare delivery.

Tahaei et al. work conducts a systematic review to explore ethical considerations in human-AI interaction, examining challenges and implications of integrating AI into human life. Through synthesizing existing literature, the study identifies ethical dilemmas such as privacy, bias, accountability, and transparency. It emphasizes the need for ethical guidelines and frameworks in AI development, addressing user concerns and issues in AI technology reviews.

Qian Yuzhou et al. (2024) examines the societal impacts of AI, including employment, privacy, security, and social structures. Through a comprehensive review of existing literature, the study identifies multifaceted impacts of AI on society, such as changes in employment patterns, privacy concerns, ethical dilemmas, and power dynamics shifts. It underscores the necessity for proactive policies and regulations to address emerging challenges, contributing to understanding user concerns in human-AI interactions.

These studies provide valuable insights into the complexities of human-AI interaction and the broader societal implications of AI technology. Understanding these dynamics is essential for addressing user concerns, fostering trust, and ensuring responsible deployment of AI technologies across various domains.

## METHOD

The research methodology involved a multi-step process for collecting and analyzing 55,968 online reviews from G2.com, ProductHunt.com, and Trustpilot.com. These sites were chosen for their comprehensive and diverse user reviews on AI software, focusing on user challenges and benefits.

A lexical approach was used, employing the WordNet dictionary and the Natural Language Toolkit (NLTK) library. The process included identifying the part of speech for each word, removing stop words and duplicates, and lemmatizing the remaining words. Synonyms and antonyms were excluded to ensure accuracy. This resulted in a dictionary of 13,522 adjectives and nouns, capturing key user terms and descriptors.

The dictionary was used to rank the words in the dataset, creating a 55,968×13,522 matrix where each review was a row and each word a column, marked by binary values. Low variance words were excluded, refining the list to 586 words, providing a detailed view of user experiences.

Exploratory Factor Analysis (EFA) was conducted using SAS software, with Varimax rotation and the Unweighted Least Squares method, to identify underlying factors. The analysis determined 32 factors, later refined to 15 based on their top values, excluding those below a 0.3 threshold.

Upon obtaining the factors, the focus shifted to identifying common themes in human-AI interaction within each factor. The specific words associated with each factor showed how strongly they were related. Since these words came from the original review texts, they reflected the language and expressions users used to describe their interactions and experiences with various AI tools. The prominence of individual words within each factor provided insights into their importance and significance in the user reviews. By examining the reviews containing words related to each factor, we could identify common themes and topics of discussion, highlighting the challenges users faced in their interactions with AI software.

## PRELIMINARY RESULTS

In our initial analysis, we conducted content analysis to identify common themes within each factor. By examining the highest value words within each factor, we could determine the most significant words representing each theme. Each factor contained 586 related words pointing to a common theme, indicating consistent user experiences and challenges in their interactions with AI software. Table 1 presents the highest value words for each of the 15 factors and the corresponding themes assigned to them. This preliminary analysis provides a foundation for understanding how each theme impacts human-AI interaction. In the following sections, we delve deeper into how these identified themes influence user experiences and interactions with AI systems.

**Table 1.** factor loading in exploring challenge of Human-AI Interaction.

| Factor # | Loaded Factor | Human-AI Challenge |
|---|---|---|
| 1 | suite (0.65), ticket (0.41) | Customer Service Automation |
| 2 | plagiarism (0.63), checker (0.59) | Content Quality |
| 3 | posit (0.61), statistical (0.44) | Visualization |
| 4 | accounting (0.59) | financial and accounting operations |
| 5 | grammar (0.59), spelling (0.41) | spelling and grammar checking features |
| 6 | defence (0.59), cisco (0.42) | cloud firewall features |
| 7 | rhombus (0.57) | Security and Surveillance Technologies |
| 8 | anaconda (0.51), python (0.48) | Technical integration challenges and resource demands |
| 9 | protection (0.47), threat (0.41) | Cybersecurity features and services |
| 10 | hybrid (0.45) | Adaptive User Interface Design |
| 11 | speech (0.44), text (0.42) | Accuracy |
| 12 | data (0.37), machine (0.34) | Predictive Analytics |
| 13 | drift (0.37) | Optimizing marketing and customer engagement |
| 14 | ai (0.36), content (0.33) | Ethical Ai Decision Making |
| 15 | mathematical (0.34) | Signal Integrity |

## Customer Service Automation

In customer service automation, companies use technology, such as AI-powered chatbots, to handle customer inquiries and issues automatically. While this can be efficient for simple questions, it presents challenges when faced with complex issues requiring empathy and understanding.

For instance, imagine a customer contacts a company about a billing discrepancy or a technical problem with a product. Instead of speaking to a human representative, they're directed to an AI chatbot. However, the chatbot offers generic responses and fails to grasp the nuances of the problem. Despite the customer's attempts to explain, the chatbot continues to provide scripted solutions that don't address the root cause of the issue.

This scenario highlights a key challenge of customer service automation: the inability of AI systems to empathize with customers and understand complex issues. While automation can streamline processes and handle routine inquiries efficiently, it lacks the human touch necessary for resolving intricate problems effectively. As a result, customers may experience frustration and dissatisfaction, potentially harming the company's reputation.

In essence, the challenge of customer service automation lies in finding a balance between technological efficiency and human empathy to ensure satisfactory customer experiences.

Some related words in this factor are Understanding, satisfaction, Frustrating, Challenge, Inability.

## Content Quality

Ensuring top-notch content quality in human-AI interaction presents a multifaceted challenge, crucial for delivering accurate and engaging information. From maintaining grammatical accuracy and language nuance to detecting plagiarism and verifying information integrity, various hurdles abound. While AI systems can assist in content generation, they often falter with language subtleties, fact-checking, and contextual understanding, key components for producing high-quality content. Thus, achieving optimal content quality necessitates striking a delicate balance between AI automation and human expertise to effectively address these complex challenges.

In a hypothetical scenario, envision an online news platform employing AI algorithms to automatically generate articles on trending topics. Despite its ability to churn out content swiftly, the AI system grapples with upholding content quality. For instance, an article discussing a recent scientific breakthrough is published with inaccuracies and lacks proper source citations, leading to the dissemination of misinformation. Despite efforts to enhance the AI's language processing capabilities, it continues to struggle with comprehending scientific terminology and context nuances. Consequently, the platform faces backlash for circulating unreliable information, damaging its reputation and eroding trust among readers. This scenario underscores the significant challenge of preserving content quality in human-AI interaction, underscoring the imperative for human oversight and intervention to ensure the accuracy and credibility of automated content generation processes.

Some related words in this factor are plagiarism, grammar, spelling, accuracy, quality, integrity, clarity, premium, professional, insightful.

## Visualization

visualization in human-AI interaction lies in the ability of AI systems to accurately interpret and present data in a visual format that is easily understandable and meaningful to humans. While AI algorithms can generate visualizations quickly, they may struggle with creating insightful and effective visual representations due to limitations in understanding context, relevance, and user preferences. Additionally, ensuring the accuracy and integrity of the data being visualized poses another challenge, as errors or inaccuracies in the underlying data can lead to misleading or incorrect conclusions. Therefore, achieving effective visualization in human-AI interaction requires addressing these issues to enhance the interpretability and usefulness of visualized information.

A data analytics firm employs AI for visualizing a client's marketing data, the AI system generates confusing and inaccurate visualizations. Despite attempts to improve, the AI struggles with understanding industry-specific metrics and client preferences. Consequently, the client faces challenges in interpreting the data accurately, leading to frustration and a hindrance in decision-making. This highlights the difficulty in achieving effective visualizations in human-AI interaction, emphasizing the need for human oversight to ensure clarity and accuracy in data representation.

Some related words in this factor are Visualization, Inaccurate, Clarity, Accuracy, Data, Performing.

### Financial and Accounting Operations

In a financial institution's adoption of AI for loan approvals, the system's reliance on statistical patterns leads to discriminatory outcomes and an inability to adapt to economic shifts, risking regulatory scrutiny and customer trust.

The challenge of financial interaction between humans and AI involves tasks related to managing finances, such as accounting, payroll processing, tax calculation, transaction tracking, and invoice management. These tasks require accuracy, timeliness, and compliance with financial regulations. However, integrating AI into financial processes can present various challenges, including ensuring the security and integrity of financial data, streamlining complex financial operations, and adapting to changing financial landscapes. Additionally, there may be concerns about the reliability and accuracy of AI algorithms in handling financial tasks, as errors or inaccuracies could have significant financial implications. Overall, navigating the intersection of finance and AI involves addressing these challenges to effectively leverage AI technology while maintaining financial stability and trust.

Some related words in this factor are accounting, sage, financial, invoice, payroll, tax, transaction, payment, tracking, banking, recurring, costing, revenue, audit, budget, monthly, income, asset.

### Spelling and Grammar Checking Features

The challenge of typing in human-AI interaction refers to difficulties or issues that arise when users input text into AI systems. This could include problems such as typos, misspellings, or errors in input, which can impact the accuracy or effectiveness of the AI's response or output. Additionally, typing challenges may involve the need for users to input text quickly and accurately, especially in real-time communication or data entry scenarios. Improving typing efficiency and accuracy can enhance the overall interaction and user experience with AI systems.

Some related words in this factor are Typing, Spelling, Checker, Sentence, Content, Editing, Grammar, Text.

### Cloud Firewall Features

Human interaction with AI in cloud firewall solutions, such as Cisco Secure Firewall Threat Defense Virtual, can present several challenges. These systems often require significant expertise to deploy, configure, and manage effectively. Users might struggle with complex interfaces, such as command-line consoles or sophisticated graphical user interfaces, which can be daunting without proper training. Additionally, frequent updates and the need to understand advanced security features can overwhelm IT staff. The AI-driven automation intended to ease management can sometimes lead to confusion or mistrust, especially if the system's decision-making process is not transparent.

Balancing the advanced capabilities of AI with user-friendly operation and clear communication remains a key challenge in these interactions.

Some related words in this factor are clarity, insight, representation, visibility, network.

### Security and Surveillance Technologies

AI-enhanced surveillance systems, such as those integrating facial recognition, real-time alerting, and behavior analysis, offer unprecedented levels of security. They enable swift identification and response to potential threats, thereby enhancing safety and operational efficiency. However, these systems also raise significant concerns regarding privacy, data security, and ethical use. The ability of AI to process and analyze vast amounts of personal data can lead to unauthorized surveillance, data breaches, and misuse of sensitive information. Ensuring robust security measures, such as data encryption, access controls, and compliance with privacy regulations like GDPR, is essential to mitigate these risks. Additionally, transparent policies and ethical guidelines must be established to maintain public trust and ensure that AI-driven surveillance is used responsibly and fairly. Balancing the benefits of advanced security features with the imperative of protecting individual privacy remains a critical challenge in the deployment of AI in security and surveillance contexts.

Some related words in this factor are surveillance, security, alert, monitoring, privacy, threat, visible.

### Technical Integration Challenges and Resource Demands

Technical integration challenges and resource demands are significant hurdles in the deployment of sophisticated AI systems. Integrating AI technologies into existing infrastructures often involves complex configurations and compatibility issues, requiring substantial technical expertise and time. Resource demands, such as high memory and processing power, can strain existing systems, leading to performance bottlenecks and inefficiencies. These challenges can delay implementation timelines and increase costs, necessitating careful planning and robust support systems. Addressing these issues is crucial to ensure seamless integration, optimal performance, and the long-term sustainability of AI-driven solutions.

Some related words in this factor are Implement, collaborative, Integration, compatible, adoption, complicated, effective.

### Cybersecurity Features and Services

Integrating cybersecurity features and services into artificial intelligence (AI) systems presents several unique challenges. As AI systems increasingly rely on vast amounts of data to function effectively, ensuring the security and privacy of this data becomes paramount. Cyber threats targeting AI can exploit vulnerabilities in the data pipeline, such as data poisoning, where malicious actors introduce corrupted data to skew AI outputs. Additionally, AI models themselves can be susceptible to adversarial attacks, where inputs are subtly manipulated to produce incorrect results. Implementing robust cybersecurity measures requires advanced encryption techniques, secure access controls, and continuous monitoring to detect and mitigate these threats. Furthermore, the complexity of AI systems can make it difficult to ensure comprehensive protection without impacting performance. Compliance with regulations such as GDPR and CCPA adds another layer of complexity, as organizations must balance the need for data-driven AI innovation with stringent data privacy requirements. Addressing these challenges is crucial to maintaining the integrity, reliability, and trustworthiness of AI systems in an increasingly interconnected digital landscape.

Some related words in this factor are Protection, Threat, Malicious, Privacy, Visibility, Monitoring, Data.

### Adaptive User Interface Design

Adaptive User Interface Design enables a more personalized, efficient, and accessible user experience, driven by AI technologies that analyze user data and adapt the interface in real-time. One of the significant issues is the potential for over-reliance on AI algorithms to make interface decisions, which can lead to unintended consequences or biases in the user experience. Additionally, there may be concerns regarding user privacy and data security, as AI algorithms often require access to large amounts of user data to personalize interfaces effectively. Moreover, ensuring that AI-driven interfaces remain intuitive and user-friendly can be challenging, as complex AI models may produce unpredictable or confusing outcomes for users. Striking the right balance between AI-driven personalization and maintaining user control and transparency is crucial to addressing these challenges and creating successful adaptive user interfaces.

Some related words in this factor are Dependency, Legacy, Overwhelming, privacy, integration, adaptable.

### Accuracy

In the realm of human-AI interaction, one significant challenge is ensuring "accuracy." It encompasses the need for AI systems to generate precise and reliable outputs across various tasks, including speech recognition, text transcription, and data processing. The accuracy of AI algorithms directly impacts the effectiveness and usability of applications, as users rely on these systems to provide correct information and insights. Moreover, accuracy is essential for building trust and confidence in AI technologies, as users expect consistent and dependable performance in their interactions with AI-powered interfaces and services. Therefore, addressing the challenge of accuracy is critical for enhancing the overall quality and user experience of human-AI interaction.

Some related words in this factor are speech, inaccurate, text, transcription, accuracy, processing, machine, quality, editing, express, automatic, facial, artificial, meaningful, semantic, wrong, precise, valuable, content.

### Predictive Analytics

Predictive analytics, while incredibly powerful, can present several challenges in human-AI interaction. One primary issue is the potential for biased or inaccurate predictions. If the data used to train predictive models is skewed or incomplete, it can lead to biased outcomes, perpetuating existing inequalities or making flawed decisions. Moreover, reliance on predictive analytics may also raise concerns about privacy and data security, especially if sensitive personal information is used without proper consent or protection. Additionally, there's the risk of over-reliance on automated predictions, leading to a reduction in critical thinking and human judgment, potentially overlooking important contextual factors that cannot be captured by algorithms alone. Therefore, while predictive analytics can offer valuable insights, it's crucial to address these issues to ensure responsible and effective human-AI interaction.

Some related words in this factor are Predictive, Analytics, Data, Mining, Statistical, Unstructured, Processing, Visualization, Preparation, Warehouse, Quality, Parallel, Privacy, Representation, Historical, Accuracy, Model, Insight, Dynamic, Scalable, Effective, Flexibility, Solution, Pattern, Algorithm, Regression, Decision, Risk, Behavior, Real-time, Optimization, Reliability, Machine-learning, Forecasting, Clustering, Anomaly, Detection, Prescriptive, Intelligence, Strategy, Event, Horizon, Reporting, Monitoring, Planning, Business.

### Optimizing Marketing and Customer Engagement

One key issue in optimizing marketing and customer engagement through AI is the potential for unintended bias in decision-making algorithms. AI systems learn from historical data, which may contain biases based on factors like race, gender, or socioeconomic status. If these biases are not identified and addressed, AI algorithms can perpetuate and even exacerbate existing inequalities. This can lead to unfair treatment of certain groups of customers and damage a company's reputation. Therefore, mitigating bias in AI algorithms is critical to ensuring fair and equitable marketing and customer engagement practices.

Some related words in this factor are Customer, data, historical, Bias, Optimization, Marketing, Algorithm, Unintended, Inequality, Fair, Equitable, Reputation, Mitigating, Treatment, Unfair.

### Ethical AI Decision-Making

In the realm of human-AI interaction, ethical decision-making by AI systems poses a significant challenge. AI algorithms often make decisions based on complex calculations and patterns learned from data. However, these decisions can sometimes raise ethical concerns, such as fairness, transparency, and accountability. For instance, in hiring processes, AI algorithms might inadvertently discriminate against certain demographics or reinforce existing biases present in historical data. Additionally, AI systems may lack the ability to explain their decisions comprehensibly, leading to a lack of transparency and trust among users. Addressing these ethical challenges requires careful consideration of the societal implications of AI decision-making and the development of frameworks and guidelines to ensure that AI systems prioritize fairness, transparency, and accountability in their operations.

Some related words in this factor are Ethical, Decision-making, Transparency, Accountability, Fairness, Discrimination, Bias, Trust, Frameworks, Guidelines.

### Signal Integrity

The challenge of signal in human-AI interaction revolves around the accurate transmission and interpretation of information between humans and AI systems. Signals are essential for communication between these entities, conveying commands, data, or feedback. However, ensuring the clarity and reliability of signals can be challenging, particularly in environments with high noise levels or complex data streams. For instance, in a scenario where AI systems rely on sensor data to make decisions, signal interference or distortion could lead to erroneous interpretations and subsequent actions. Additionally, variations in signal strength or frequency may hinder the effectiveness of communication between humans and AI, resulting in misunderstandings or incomplete information exchange. Addressing the challenge of signal requires implementing robust signal processing techniques, optimizing communication protocols, and developing adaptive AI systems capable of interpreting signals accurately in diverse contexts. Ultimately, overcoming signal-related challenges is essential for facilitating seamless and efficient interactions between humans and AI.

Some related words in this factor are Signal, Simulation, Processing, Data, Power, High, Industrial, Precision, Machine, Statistical, Performing, Dynamic, Predictive, Network, Scalable, Volume, Repository, Scalability, Mining, Transaction, Payment, Quality, Accuracy, Scale, Surveillance, Visibility.

## FUTURE RESEARCH

In the next step, each review will be delved into more deeply, and all the words within each factor will be analyzed. This detailed examination is aimed at developing a comprehensive summary for each factor. By thoroughly investigating the language and context of the user reviews, a final summary that captures the essence and nuances of each factor will be provided, further illuminating the specific challenges and themes in human-AI interactions.